\documentstyle[12pt,mont98_desy,amssymb,epsfig]{article}

\def\be{\begin{equation}}
\def\ee{\end{equation}}
\def\bea{\begin{eqnarray}}
\def\eea{\end{eqnarray}}

\begin{document}

\title{III.~~THE PERMISSIBLE EQUILIBRIUM POLARISATION DISTRIBUTION IN A STORED 
 PROTON BEAM 
\footnote{Updated version of a talk presented
at the 15th ICFA Advanced Beam Dynamics Workshop: ``Quantum
Aspects of Beam Physics'', Monterey, California, U.S.A., January 1998.
Also in DESY Report 98--096, September 1998.}
}
\author{{\underline {D.P.~BARBER}}, K.~HEINEMANN, G.H.~HOFFST\"ATTER and
M.~VOGT }

\address{Deutsches Elektronen--Synchrotron, DESY, \\
22603 Hamburg, Germany.\\E-mail: mpybar,~heineman,~hoff,~vogtm~@mail.desy.de}


\maketitle
\abstracts{ We illustrate the use of the invariant spin field for describing
permissible equilibrium spin distributions in high energy spin polarised proton
beams.}

\section{A problem and its solution}
Following the successful attainment of longitudinal $e^{\pm}$  polarisation 
in HERA (Article II) it is natural to consider whether it would be
possible to complement the polarised $e^{\pm}$ with $820 GeV$ polarised 
protons \cite{erice95,krischcrap}.

As pointed out in Article I, a stored polarised proton beam can only
be obtained by injecting and then accelerating a prepolarised beam provided
by a suitable source. However, I comment on another concept in the Appendix.

A major obstacle to reaching high energy with the polarisation intact is 
that the spins must negotiate groups of spin--orbit resonances every $523 MeV$
(Article I) since the spin tune is approximately proportional to the energy.

However, this problem can be ameliorated by the inclusion of Siberian Snakes 
\cite{dk76,dk78}. These are magnet systems which rotate spins by $180$
 degrees
around an axis in the horizontal plane independently of the energy of the
particle. By the installation of suitable combinations of snakes, the spin 
tune $\nu_{spin}$ can be fixed at one half and then by suitable choice of
orbital tunes, resonances can be avoided at all energies, assuming that
the dependence of spin tune on synchrobeta amplitude is weak (Article I).

Tracking simulations show that even with snakes, preservation of polarisation
up to high energy is nontrivial. For example a $1$ milliradian orbit 
deflection at $820 ~GeV$ causes a $90$ degree spin 
rotation (Article I, Eq.~(4)).
 Thus one should check first
whether the spin distribution permitted by the requirement of equilibrium
at a chosen high energy would be acceptable. There would be no point in
trying to accelerate if the answer were negative. Moreover, to arrive
at an answer we have the ideal tool at hand, namely the
invariant spin field introduced in Article I. The measure for acceptability
is the deviation of $\hat n$ from ${\hat n}_0$ averaged across phase space. 
If the average deviation were, say, $60$ degrees, then even with 
$|{\vec P}_{eq}(\vec u; s)| = 1$  at each point in phase space,
a polarimeter would only record about $50 \%$ polarisation. Thus the optic and
ring layout must be chosen so that the deviation is minimised. 
The invariant spin field can be calculated using the numerical technique
`stroboscopic averaging' of the computer code SPRINT \cite{hh96}
\footnote{The new version of the SODOM algorithm \cite{yok98} gives 
equivalent results. See Article I.}. 

Examples of the invariant spin field for 
a HERA proton optic with a suitable snake layout are shown in the figures.
In this simulation the protons only execute integrable vertical betatron 
motion. Each figure shows the locus, on the surface of a sphere, of 
the tip of the $\hat n$ vector `attached' to its phase space ellipse at an
interaction point on the ring where ${\hat n}_0$ is vertical.
The parameters are shown in the captions. 
An emittance of $4 \pi$ mm mrad corresponds to `1-$\sigma$'. 
\begin{figure}[htbp]
\begin{center}
\psfig{figure=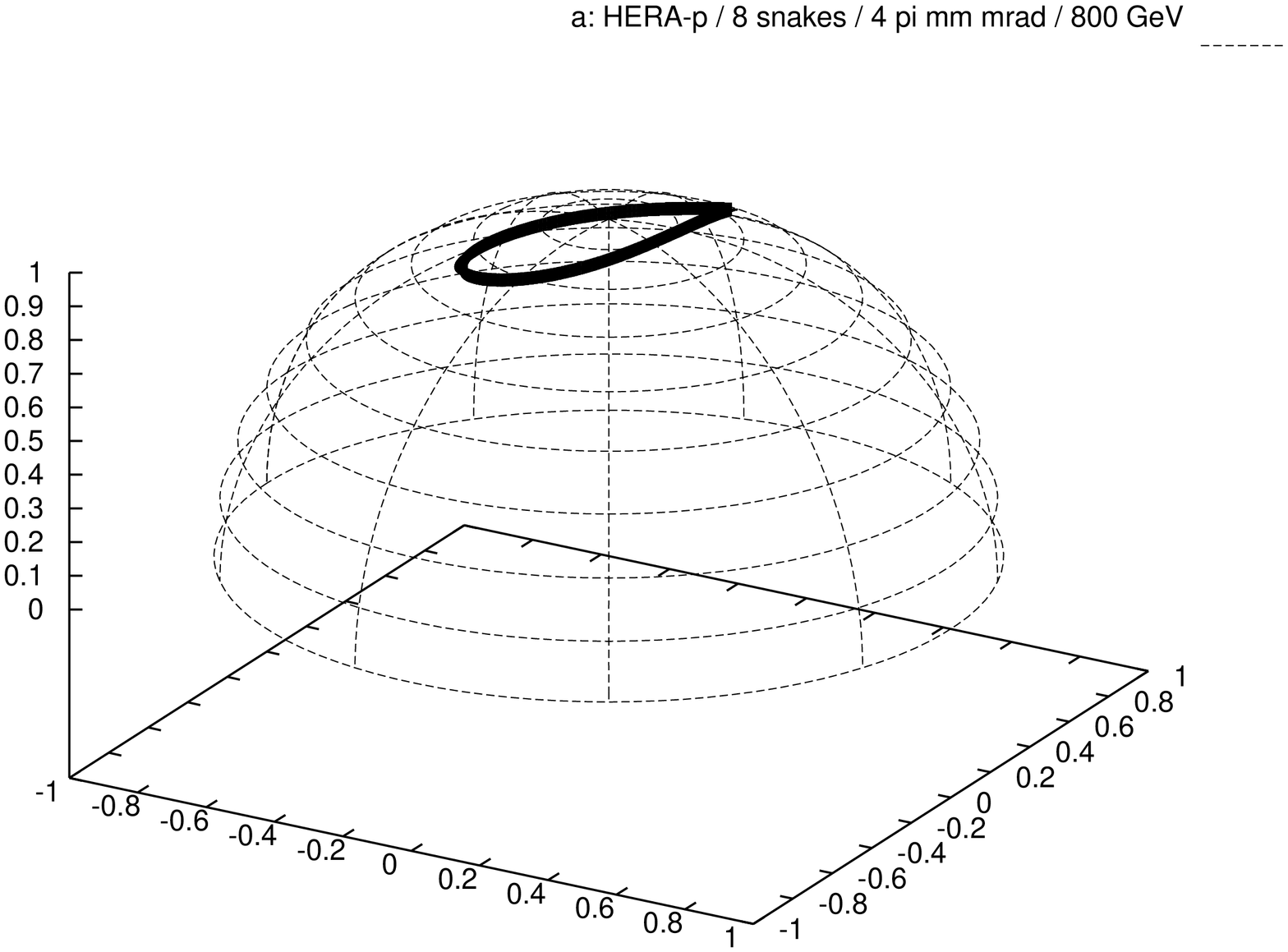,width=4.5cm,angle=-90}
\psfig{figure=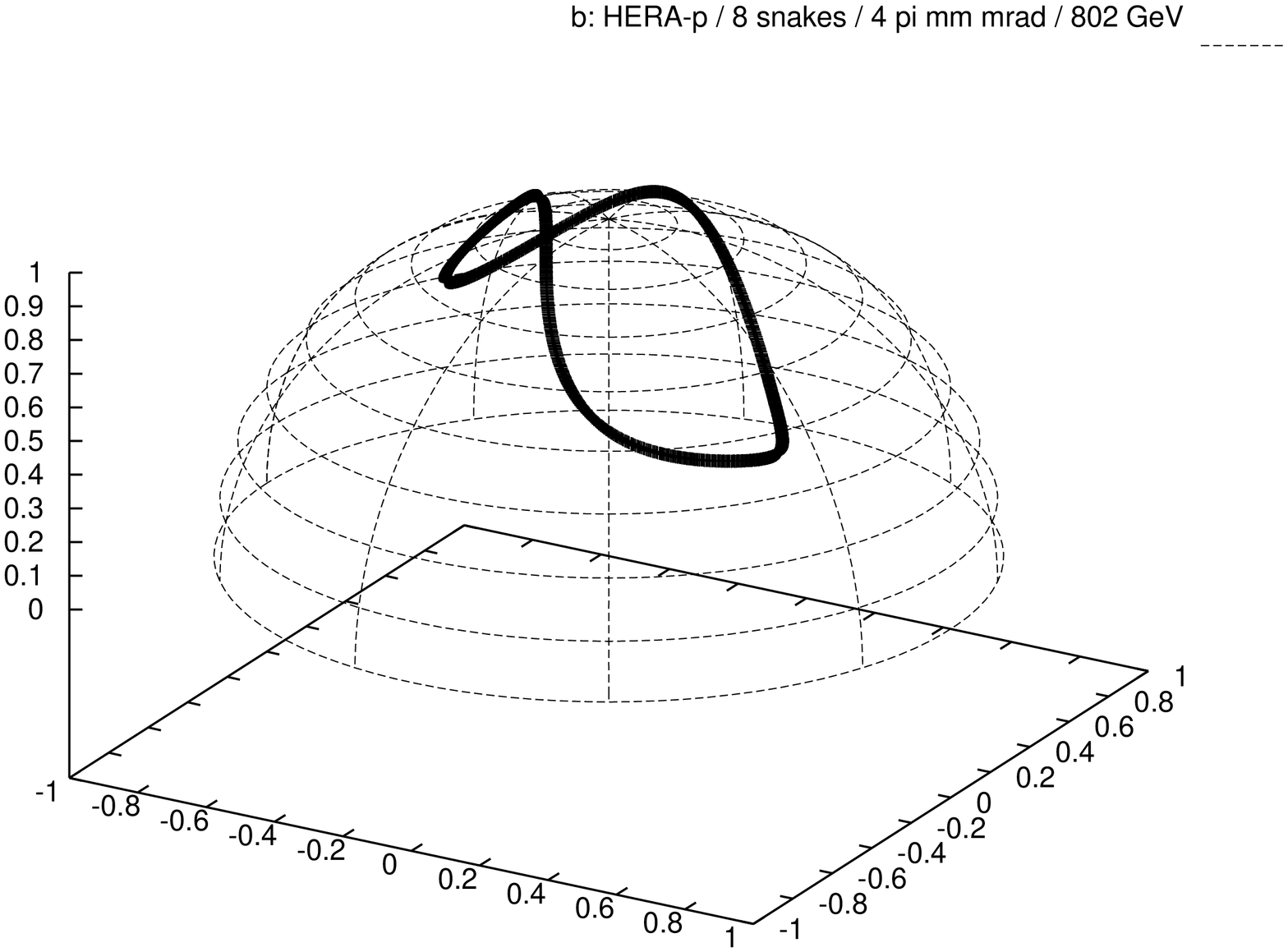,width=4.5cm,angle=-90}
\end{center}
\caption{\footnotesize
  The $\hat{n}$--vector for the $4 \pi$ mm mrad ellipse at $800~GeV$ (left)
and $802~GeV$ (right).}
\label{fg:NaxSph2}
\end{figure}
\begin{figure}[htbp]
\begin{center}
\psfig{figure=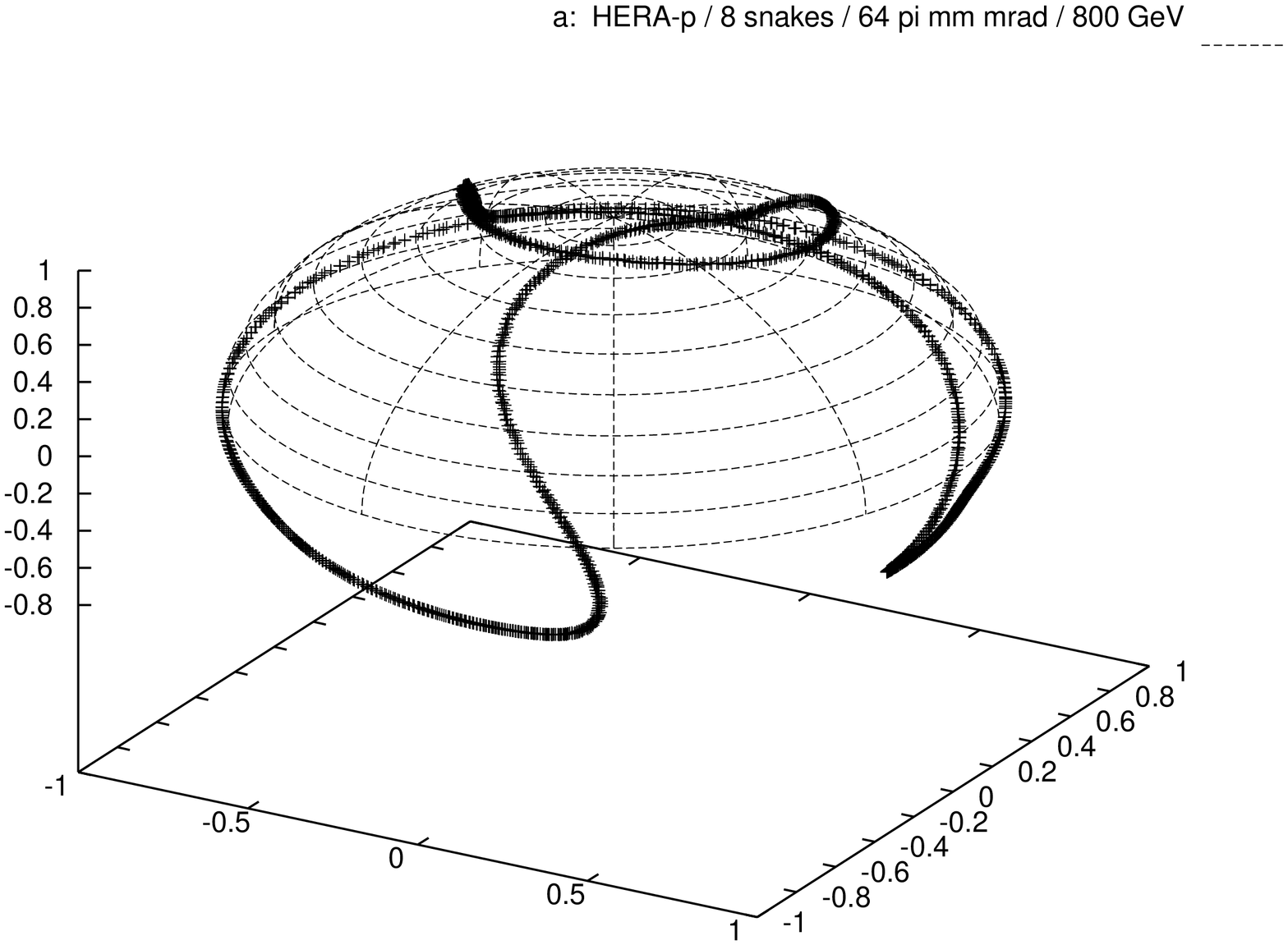,width=4.5cm,angle=-90}
\psfig{figure=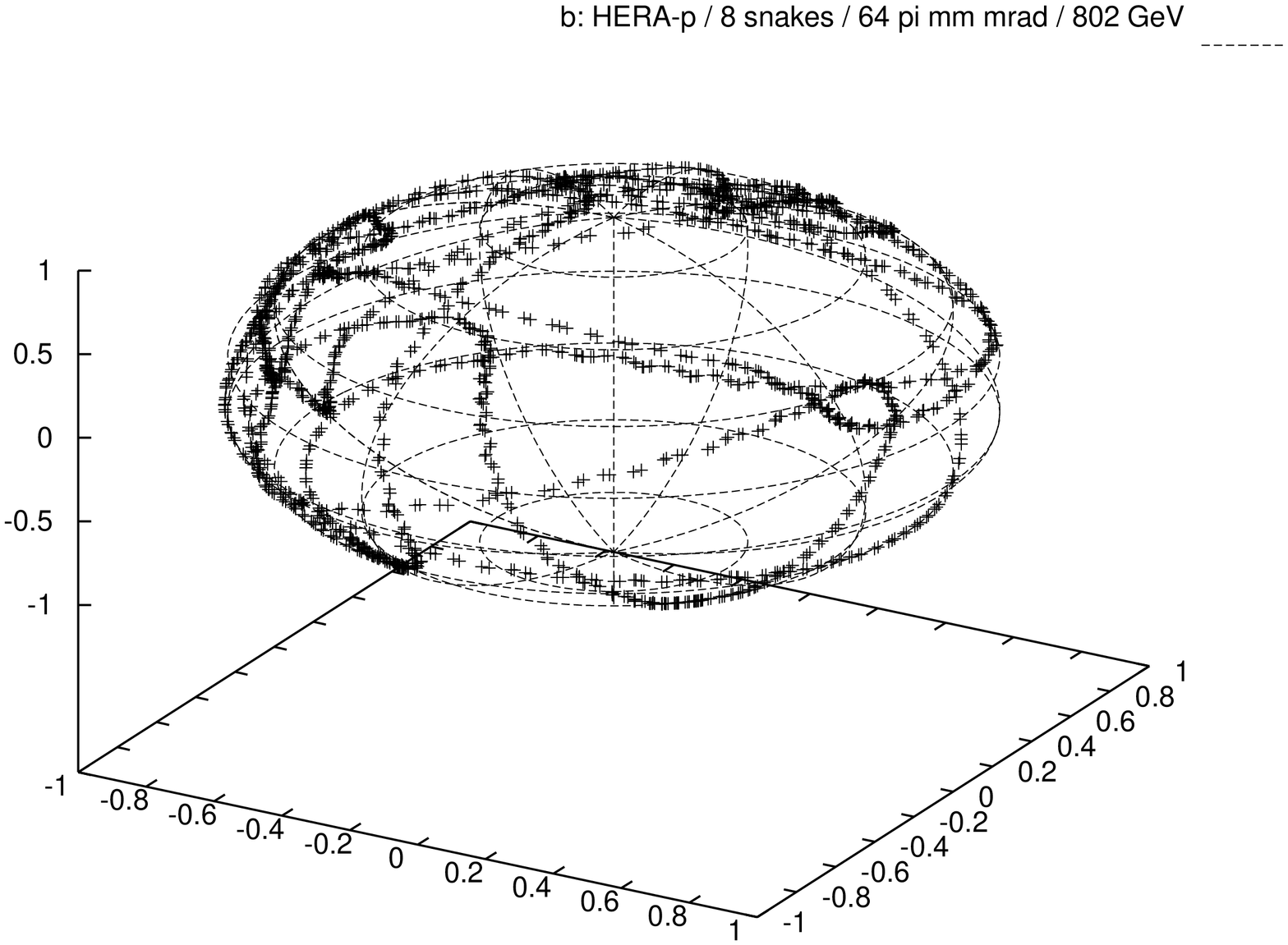,width=4.5cm,angle=-90}
\end{center}
\caption{\footnotesize
The $\hat{n}$--vector for the $64 \pi$ mm mrad ellipse at $800~GeV$ (left)
and $802~GeV$ (right).}
\label{fg:NaxSph1}
\end{figure}
The energy $800~GeV$ lies well below a resonance structure that survives
even in the presence of snakes and
$802~GeV$ is just below this structure. For particles at 1-$\sigma$ 
the spin field is well aligned at $800~GeV$. At 4-$\sigma$ it has opened
well beyond $90$ degrees at some phases.
At $802~GeV$ the 1-$\sigma$ locus deviates by more
than $30$ degrees from $\hat n_0$ at some orbital phases and at 4-$\sigma$
the field is almost isotropic!
In all four cases the locii are closed as required by the periodicity
condition ${\hat n}(\vec u; s) = {\hat n}(\vec u; s+C)$ (Article I).


A  distribution of spins aligned along an invariant spin field is 
the ideal starting point for long term tracking
studies of spin stability at fixed energy since deviations from equilibrium
are then easy to discern.

\section*{Appendix}
It has been suggested that by using Stern--Gerlach (SG) forces to drive 
coherent synchrobeta motion and thereby separate particle bunches into
`spin--up' and  `spin--down' parts, a proton beam could effectively be
polarised \cite{pust}. The scheme using transverse SG forces requires running
close to spin--orbit resonance but figure 2 illustrates that at high amplitude
spin directions become isotropic so that the SG effect would average 
away. In any case the basic scheme might fail as a result of conservation
laws \cite{derb90,bhr1,bal98}. 
The longitudinal SG effect \cite{pust} would be subject to mixing due to
synchrotron oscillation unless some very special means were found
to prevent it. Moreover, the longitudinal SG force is a total time
derivative of a function of the fields and could integrate to 
zero \cite{kh96,krip97,ptit98,derbkh}.


\section*{References}
\begin{thebibliography}{99}

\bibitem{erice95} D.P.~Barber in the proceedings of the conference
``The Spin Structure of the Nucleon'',
Erice 1995, World Scientific, April 1998.

\bibitem{krischcrap}
``Acceleration of Polarized Protons to 820 GeV at HERA'',
 University of Michigan Report UM--HE 96-20, November 1996
and references therein.
\bibitem{dk76}
Ya.S. Derbenev and A.M. Kondratenko,  Sov.Phys.Dokl., {\bf 20}, 562
(1976).

\bibitem{dk78}                                              
Ya.S. Derbenev et al.,  Particle Accelerators, {\bf 8}, 115 (1978).

\bibitem{hh96}
K.~Heinemann and G.~H.~Hoffst{\"a}tter,
Physical Review   {\bf E 54} N$^0$4, 4240 (1996).

\bibitem{yok98}
K. Yokoya, DESY report 99-006 (1999).

\bibitem{pust}
M. Pusterla, contribution to these Proceedings and references therein.

\bibitem{derb90}
Ya.S. Derbenev, University of Michigan - Ann Arbor  preprint, UM-HE 90-32 
(1990).

\bibitem{bhr1}
D.P. Barber, K. Heinemann and G. Ripken, Z.f.Physik, {\bf C64},117--167
 (1994).

\bibitem{bal98}
V.V.  Balandin and N.I. Golubeva,  DESY report  98-16 (1998).

\bibitem{kh96}
K. Heinemann, DESY report 96-229 (1996) and
 Los Alamos archive: physics/9611001.

\bibitem{krip97}
I.B. Khriplovich and A.A. Pomeransky,
J. Exp. Theor. Phys., {\bf 86}, 839 (1998) and 
Los Alamos archive: gr-qc/9710098. 


\bibitem{ptit98} 
V. Ptitsin, ``On the longitudinal Stern--Gerlach force'', unpublished  notes
1998.

\bibitem{derbkh}
Ya.S. Derbenev and K. Heinemann, private communications 1998.


\end{thebibliography}
\end{document}